\documentclass[useAMS,usenatbib]{mn2e}
\usepackage{graphicx}
\usepackage{psfrag}

\title[The braking indices of radiopulsars]{
   The influence of nondipolar magnetic field
   \\
   and neutron star precession
   \\
   on braking indices of radiopulsars. 
}
\author[D.P. Barsukov and A.I. Tsygan]{
D.P. Barsukov$^{1}$
and A.I.Tsygan $^{1}$\thanks{E-mail:tsygan@astro.ioffe.ru} 
\\
$^{1}$
Ioffe Physical-Technical Institute of the Russian Academy of Sciences,
Saint-Petersburg, Russia
}
\begin{document}

%

\maketitle

\label{firstpage}

\begin{abstract}
Some of radiopulsars have anomalous braking index values 
$n = \Omega \ddot{\Omega} / \dot{\Omega}^{2} 
   \sim \pm ( 10^{3} \div 10^{4} )
$.
It is shown that such anomalous values may be related to 
nondipolar magnetic field. 
Precession of a neutron star leads to rotation 
(in the reference frame of the star)
of its angular velocity $\vec{\Omega}$ around
the direction of neutron star magnetic dipole moment $\vec{m}$
with an angular velocity $\vec{\Omega}_{p}$.
This process may cause the altering of electric current flowing 
through the inner gap and 
consequently the current losses 
on the time scale of precession period $T_{p} = 2\pi / \Omega_{p}$.
It occurs because the electric current in the inner gap 
is determined by Goldreich-Julian charge density
$\rho_{GJ} = -\frac{ \vec{\Omega} \cdot \vec{B} }{ 2\pi c }$,
that depends on the angle between direction of 
small scale magnetic field and the angular velocity $\vec{\Omega}$.

\end{abstract}


\section{Introduction}
Radio pulsars have been discovered more than 40 years ago 
\cite{Hewish1968}
and at present many thousand papers are devoted to these objects.
Despite of large progress in the understanding 
of processes in pulsar magnetospheres 
still many important questions are unclear.
One of them is the value of pulsar braking indices.
If a pulsar were just a simple magnet 
with dipolar magnetic momentum $\vec{m}$ 
rotating with angular velocity $\vec{\Omega}$,
then the braking index $n = \ddot{\Omega}\Omega / \dot{\Omega}^{2}$
would be equal to $3$.
The taking into account of evolution of 
angle $\chi$ between vectors $\vec{m}$ and $\vec{\Omega}$ 
yields
$n = 3 + 2cot\chi$
\cite{DavisGoldstein1970,Beskin_book}. 
Observations of some pulsars are well within theoretical predictions.
For example, the braking index of pulsar Crab is equal to 2.5
and pulsar Vela has $n\approx 1.4$
\cite{Beskin_book}.
So it seems that more sophisticated magnetospheric models 
will be able to remove this discrepancy,
see for example
\cite{Melatos1997,Timokhin2005,Timokhin2006,Contopoulos2006,Contopoulos2007,
Timokhin2007_Force_free,Timokhin2007_Differentially}.
However, many isolated pulsars
have very large positive or, sometimes, negative braking indices
upto $|n| \sim 10^{4}$.
Sometimes such large values may be due to unobserved glitches 
or timing noise \cite{Alpar2006}.
For example, in \cite{Alpar2006} it is shown that
all negative braking indices may be associated with  
unresolved glitches
which has occurred between intervals of observations of the pulsars.
But in some cases refined measurements of braking indices 
show that at least some pulsars 
have large and sometimes negative braking indices
$n \sim \pm (10-10^{2})$ \cite{Galloway1999}.
For example, the pulsars B0656+14 and B1915+13 have braking indices  
$n \approx 14.1 \pm 1.4$ and $n \approx 36.08 \pm 0.48$,
correspondingly,
while pulsar B2000+32 and B1719-37 have very large negative 
braking indices $n \approx -226 \pm 4.5$
and $n \approx -183 \pm 10$, correspondingly
\cite{Galloway1999}.
There are some explanations of such braking index values.
For example, the values like $n \sim \pm 10$ may be related to  
nonstandard mechanisms of pulsar braking,
like neutron star slowing down
due to neutrino emission \cite{Peng1982} 
or due to interaction with circumpulsar disk 
\cite{Menou2001,Malov2004,Li2006},
see \cite{Malov2001} for review of possible mechanisms 
and its comparison with observations.
The large braking indices may be explained by
the rapid changing of magnetic field.
Such changes may be caused by Hall-drift instabilities
\cite{Geppert2002_Hall_drift,Geppert2002_Non_linear,Geppert2007}.
Also the positive braking indices of old pulsars 
may be explained by the relaxation of angular velocities of
neutron star crust and superfluid between two glitches 
\cite{Alpar2006}.
The large braking indices of some pulsars 
may be related to Tkachenko waves \cite{Popov2008}.
In this paper we present a some model based on 
works \cite{Biryukov2006} and \cite{Contopoulos2007_Note_on_cyclic},
see also \cite{Biryukov2007,Link2006_Strong,Geppert2007},
where it has been shown that large braking indices
may be explained by the existence of some internal cyclic process 
in a pulsar.

\section{The torque}
There are many mechanisms, 
that may be responsible for pulsar braking
\cite{Beskin_book,Malov2001}.
Let us shortly describe two of them.
The first of them was proposed before pulsar discovery
by \cite{Pachini1967}.
According to this mechanism the rotation energy $E$ 
and the rotation momentum $\vec{M}$ of a neutron star 
are carried away by magnetic dipole radiation,
that, of course, leads to a torque acting on the star.
Because of the presence of currents and charges in pulsar magnetosphere, 
calculation of the strength of this torque is 
a highly complicated 
and, at present, uncompleted task,
see for example 
\cite{Timokhin2006,Timokhin2007_Force_free,Spitkovsky2006,Beskin_book}.
However, the torque  can be estimated within a pure vacuum model 
of pulsar magnetosphere, where magnetospheric currents and charges 
are absent \cite{Deutch1955}. 
In this case, the rotation torque $\vec{K}_{dip}$ acting on a neutron star, 
can be written as
\cite{DavisGoldstein1970,Melatos2000}:
\begin{equation}
\vec{K}_{dip} = K_{0} \left(
                                \vec{e}_{m} \cos\chi
                              - \vec{e}_{\Omega}
                              + R_{dip} [ \vec{e}_{\Omega} \times \vec{e}_{m} ]
                      \right)
\label{Kdip_def}
,
\end{equation}
where $\vec{\Omega}=\Omega \vec{e}_{\Omega}$ 
is the angular velocity of the star,
$\vec{m} = m \vec{e}_{m}$ is the dipolar magnetic momentum of the star,
$\vec{e}_{\Omega}$ and $\vec{e}_{m}$ are
two unit vectors, directed, correspondingly, along 
$\vec{\Omega}$ and $\vec{m}$,
$\chi$ is the angle between $\vec{e}_{\Omega}$ and $\vec{e}_{m}$.
In the case of a small rotation speed $\Omega a / c  \ll 1$,
where $a\approx 10 {\mathrm km}$ is the radius of the star,
$K_{0}$ and $R_{dip}$ can be written as
\cite{DavisGoldstein1970}
\begin{equation}
K_{0} = \frac{2}{3} \frac{ m^{2} \Omega^{3} }{ c^{3} }
\label{K0_def}
,
\end{equation}
\begin{equation}
R_{dip} = \frac{3}{2} \xi 
          \left( \frac{c}{ \Omega a } \right)
          \cos\chi
\label{Rdip_def}
.
\end{equation}
In \cite{DavisGoldstein1970,Goldreich1970} the coefficient $\xi$ is taken 
to be equal to $1$.
We guess that the value $\xi= 3 / 5$ adopted in \cite{Melatos2000}
may be closer to reality, 
because it corresponds to absence of electric current sheet
on the neutron star surface. 
Thus, we will further consider $\xi= 3 / 5$.
Expression (\ref{Kdip_def}) 
is valid at any values of $\Omega a / c$,
although coefficients $K_{0}$ and $R_{dip}$ slightly 
distinguish from their nonrelativistic values, 
defined by (\ref{K0_def}) and (\ref{Rdip_def}),
\cite{Melatos2000}.

The second mechanism (current losses) is due to the electric current 
$\vec{j}$, that flows along open field lines,
see for example \cite{Beskin1984,Beskin_book,Beskin_Radiopulsars,Beskin2006}.
This current flows from light cylinder, crosses pulsar polar cap diode,
intersects the star surface, and 
goes into deep crustal layers.
After this it begins to move to surface and 
then transforms to the backward current
flowing along the separatrix between open and closed field lines.
It is worth to note that when the electric current travels inside the crust
it sometimes flows across the magnetic field, 
that leads to torque $\vec{K}_{cur}$ acting on the star polar cap.
The strength of this torque is calculated in 
\cite{Jones1976} as:
\begin{equation}
\vec{K}_{cur} = - K_{0} \cdot 
                  \alpha \cdot
                  \vec{e}_{m} \cos\chi
\label{Kcur_def}
,
\end{equation}
where parameter $\alpha( \Omega, \chi, \phi )$ 
characterizes the electric current value
\cite{Beskin2006,Beskin_book}.
In the case of a circular pulsar tube cross section 
the parameter can be estimated as
\cite{Beskin_book,Beskin2006,Tsygan2009}
\begin{equation}
\alpha = \frac{3}{4\cos\chi} 
                     \left(
                            \frac{ j_{N} }{ j_{GJ}^{0} } 
                            \left( \frac{ S_{N}(\eta) }{ S_{0}(\eta) } \right)
                           +\frac{ j_{S} }{ j_{GJ}^{0} } 
                            \left( \frac{ S_{S}(\eta) }{ S_{0}(\eta) } \right)
                     \right)
\label{alpha_def}
,
\end{equation}
where $j_{N}$ is the density of the electric current, 
that flows in the northern pulsar tube,
and $j_{S}$ is the density of electric current that flows
inside the southern pulsar tube,
$j_{GJ}^{0}\cos\chi = \frac{ \Omega B_{0} }{ 2\pi } \cos\chi $ 
is the Goldreich-Julian current,
$B_{0} = 2 m / a^{3} $ 
is the magnetic field strength at the magnetic pole,
$S_{N}(\eta)$ and $S_{S}(\eta)$ 
are the areas of the northern and southern pulsar tubes,
$S_{0}(\eta)=\pi a^{2} \left( \Omega a / c  \right) \eta^{3} $,
$\eta = r/a$, 
$r$ is the distance from the center of the star.

Following \cite{Jones1976,Qiao2001,Gil2003_Braking},
we assume that 
the star can slow down by both mechanisms simultaneously
and that the resulting torque $\vec{K}$ can be described 
just as a sum of 
partial torques $\vec{K}_{dip}$ and $\vec{K}_{cur}$: 
\begin{equation}
\vec{K} = \vec{K}_{dip} + \vec{K}_{cur}
\label{Ksum_def}
.
\end{equation}
Thus, the equation of rotation momentum loss can be written as
\begin{equation}
\frac{ d\vec{M} }{ dt } = \vec{K}_{dip} + \vec{K}_{cur}
\label{eqn_M_balance}
.
\end{equation}

\section{Spherical symmetry case}
At first case we assume that a neutron star is an absolutely rigid sphere.
Particularly, we neglect any star deformations 
and any viscosity and dissipation inside the star.
Consequently, we can suppose that the star rotation momentum
$\vec{M} = I \vec{\Omega}$, 
where $I$ is the momentum of inertia of the star.
Under this assumption equation (\ref{eqn_M_balance}) 
can be rewritten as
\begin{eqnarray}
I \frac{ d\vec{\Omega} }{ dt } = \vec{K}
              = K_{0} \cdot 
                     (
                           & \,
                             \vec{e}_{m} 
                             \left(  1 - \alpha( \Omega, \chi , \phi )  \right)
                             \cos\chi
\nonumber
\\
                           \, &
                           - \vec{e}_{\Omega}
                           + R_{dip}
                             [ \vec{e}_{\Omega} \times \vec{e}_{m} ] 
                      )
\label{Eqn_vec_domega_spher} 
.
\end{eqnarray}
Now let us to introduce three orthogonal unit vectors 
$\vec{e}_{x}$, $\vec{e}_{y}$ and $\vec{e}_{z}$
which rotate together with the star.
As vector $\vec{e}_{z}$ we use $\vec{e}_{m}$, 
direction of vectors $\vec{e}_{x}$ and $\vec{e}_{y}$ 
may be arbitrary.
Consequently, we have
\begin{equation}
\frac{ d\vec{e}_{z} }{ dt } = \left[ \vec{\Omega} \times \vec{e}_{z}  \right]
\mbox{\ ,\ }
\frac{ d\vec{e}_{x} }{ dt } = \left[ \vec{\Omega} \times \vec{e}_{x}  \right]
\mbox{\ ,\ }
\frac{ d\vec{e}_{y} }{ dt } = \left[ \vec{\Omega} \times \vec{e}_{y}  \right]
\label{dot_e_m} 
\end{equation}
and at any time $t$ the following relations are valid:
\begin{equation}
(\vec{e}_{z} , \vec{e}_{x} ) = 
(\vec{e}_{z} , \vec{e}_{y} ) = 
(\vec{e}_{x} , \vec{e}_{y} ) =   0
\mbox{\ and \ }
e_{z}^{2} = e_{x}^{2} = e_{y}^{2} = 1 
.
\end{equation}
Hence, at any time $t$ we can treat these vectors 
as an orthogonal space basis.
So it is possible to write, for example
\begin{equation}
\vec{\Omega} = \Omega \left(
                              \sin\chi \cos\phi \vec{e}_{x}
                            + \sin\chi \sin\phi \vec{e}_{y}
                            + \cos\chi          \vec{e}_{m}
                      \right)
\label{vecOmega_at_basis}
.
\end{equation}
Using the last expression, the equation (\ref{Eqn_vec_domega_spher})
may be rewritten in the form
\begin{eqnarray}
\frac{ d\Omega }{ dt } & = & - \frac{\Omega}{\widetilde{\tau}} 
                               \cdot
                               \left( 
                                        \sin^{2}\chi 
                                      + \alpha(\Omega ,\chi ,\phi ) 
                                        \cdot \cos^{2}\chi 
                               \right)
\label{dOmega_eqn}
\\
\frac{ d \chi }{ dt }  & = &  -   \frac{1}{\widetilde{\tau}}
                                  \cdot
                                  \left( 
                                          1 - \alpha(\Omega ,\chi ,\phi )  
                                  \right)
                                  \cdot
                                  \sin\chi \cos\chi
\label{dchi_eqn}
\\
\frac{ d \phi }{ dt }     & = &  -\frac{1}{\widetilde{\tau}}
                                  \cdot
                                  R_{dip}
                            =    -\frac{ 2\pi }{ T_{p} }
\label{dphi_eqn} 
,
\end{eqnarray}
where 
\begin{equation}
\widetilde{\tau}  =  \frac{3}{2} \frac{ I c^{3} }{ m^{2} \Omega^{2} } 
            \approx  1.5 \cdot 10^{8} \mbox{ year } 
                         \left( \frac{ P }{ 1s }         \right)^{2}
                         \frac{1}{B_{12}^{2}}
\label{wtau_def}
,
\end{equation}
\begin{eqnarray}
T_{p} & = & \frac{2\pi}{R_{dip}} \widetilde{\tau}
        =  
        \frac{ 4\pi }{ 3 } \frac{1}{\xi \cos\chi} 
                           \left( \frac{\Omega a}{c} \right)
                           \cdot
                           \widetilde{\tau} 
\nonumber
\\
\,  & \approx &
              1.2 \cdot 10^{5} \mbox{ year} 
                         \left( \frac{1}{\xi \cos\chi} \right)
                         \left( \frac{ P }{ 1 s }         \right)
                         \frac{1}{B_{12}^{2}} 
\label{Tp_def}
,
\end{eqnarray} 
where $B_{12} = B_{0} / 10^{12} {\mathrm G}$.
%
It is worth to note that equations 
(\ref{dOmega_eqn}-\ref{dphi_eqn})
describe neutron star rotation 
in an inertial reference frame of "rigid stars"\ .
But all vectors are represented by the basis
$( \vec{e}_{x}, \, \vec{e}_{y}, \, \vec{e}_{z} )$ altering in time. 

It is easy to see that equation (\ref{dOmega_eqn}) describes 
the losses of neutron star rotation energy 
and equation (\ref{dchi_eqn}) describes 
the evolution of the inclination angle $\chi$.
In the case of $\alpha = 1$ 
the equation (\ref{dOmega_eqn}) may be rewritten as
\begin{equation}
\frac{d P }{ dt} = \frac{P}{\widetilde{\tau}}
.
\end{equation}
So in this case parameter $\widetilde{\tau}$ is equal to
two characteristic pulsar ages $\tau = P / (2\dot{P})$.
For any other values of parameter $\alpha$ 
there is no so simple interpretation of the time $\widetilde{\tau}$.
But it is possible to treat the parameter $\widetilde{\tau}$ 
just as some characteristic time 
over which large changes of the angular velocity $\Omega$ and 
the inclination angle $\chi$ occur.
The equation (\ref{dphi_eqn}) describes the precession of the neutron star 
and $T_{p}$ may be interpreted as precession period.
It is worth to note that
the period $T_{p}$ is 3-4 orders of magnitude smaller than 
the characteristic time $\widetilde{\tau}$.
So it seems that 
the neglecting of the altering of values $\Omega$ and $\chi$ over times 
compared with precession period $T_{p}$
may be a good approximation.
Consequently, if we assume 
$\alpha = \alpha( \Omega, \chi, \phi )$ 
it will be possible to write
\begin{equation}
\frac{ d\alpha }{ dt } \approx 
      \frac{ \partial \alpha }{ \partial \phi } 
      \frac{ d\phi }{ dt }
.
\end{equation}
In this case, in equation (\ref{dOmega_eqn}) only parameter $\alpha$ 
is able to change significantly over precession period 
and hence 
\begin{equation}
\frac{ d^{2} \Omega }{ dt^{2} } \approx 
      - \frac{\Omega}{\widetilde{\tau}}
        \cdot 
        \frac{ d\alpha }{ dt }
        \cdot
        \cos^{2}\chi
\end{equation}
Thus, the braking index 
$n = \ddot{\Omega} \Omega / \dot{\Omega}^{2} $
can be estimated as
\cite{GoodNg1985}
\begin{equation}
n = \frac{\ddot{\Omega} \Omega}{ \dot{\Omega}^{2} }
  \approx 
         - \widetilde{\tau} \cdot \frac{d \phi}{ dt }
                            \cdot
                            \frac{\partial \alpha }{ \partial \phi }
                            \cdot
           \frac{   
                  \cos^{2}\chi 
                }
                {
                   \left( \sin^{2}\chi  + \alpha \cos^{2}\chi \right)^{2}
                } 
\label{n_spher} 
.
\end{equation} 
If we take into account the equation (\ref{dphi_eqn}), 
expression (\ref{n_spher}) may be rewritten as
\begin{equation}
n \approx 
           R_{dip} \cdot
                   \frac{\partial \alpha }{ \partial \phi }
                   \cdot
           \frac{   
                  \cos^{2}\chi 
                }
                {
                   \left( \sin^{2}\chi  + \alpha \cos^{2}\chi \right)^{2}
                } 
\label{n_R_spher} 
. 
\end{equation}
If we assume that $\alpha \sim 1$, 
$\cos\chi \sim \sin\chi \sim 1$, 
then the braking index $n$ may be estimated as
\begin{equation}
n \sim R_{dip} \cdot \frac{\partial \alpha }{ \partial \phi }  
\label{n_R_spher_est1}
.
\end{equation}
If we take into account the equation (\ref{Rdip_def}) 
then the braking index estimation (\ref{n_R_spher}) takes the form
\begin{eqnarray}
n & \approx &
          -\frac{3 }{ 2 } \xi
                             \left( \frac{ c }{ \Omega a } \right)
                             \cdot
                             \frac{\partial \alpha }{ \partial \phi }
                             \cdot
           \frac{   
                  \cos^{2}\chi 
                }
                {
                   \left( \sin^{2}\chi  + \alpha \cos^{2}\chi \right)^{2}
                }
\nonumber
\\
\, & \sim &
        \left( \frac{ c }{ \Omega a } \right)
        \cdot
        \frac{\partial \alpha }{ \partial \phi }  
   \sim 10^{4} \left( \frac{ 1c }{ P } \right)
        \cdot
        \frac{\partial \alpha }{ \partial \phi }         
\label{n_spherical_app}
.
\end{eqnarray}
It is necessary to note that within the current paper we assume 
that the magnetic dipole moment $\vec{m}$ 
does not change significantly 
over the precession period $T_{p}$.
Particularly, the expression (\ref{n_spherical_app}) is valid 
only if the time corresponding to large changes of 
the magnetic dipole momentum $\vec{m}$
is large enough compared with the period $T_{p}$.
                
Now following \cite{Goldreich1970,Tsygan2009}  
we define the value of function 
$f(\Omega, \chi, \phi )$
averaged over precession time as
\begin{equation}
   < f >(\Omega,\chi ) = \frac{1}{2\pi} \int_{0}^{2\pi}
                                        f(\Omega,\chi,\phi)
                                        d\phi 
.
\end{equation}
Then, averaged equations (\ref{dOmega_eqn}) and (\ref{dchi_eqn})
may be written as
\begin{eqnarray}
< \frac{d\Omega}{dt} > = - \frac{\Omega}{\widetilde{\tau}} 
                           \left(
                                   \sin^{2}\chi
                                  +<\alpha > \cos^{2}\chi
                           \right)
\\
< \frac{d\chi }{ dt } > = - \frac{1}{\widetilde{\tau}}
                            \left( 
                                   1 - <\alpha >
                            \right)
                            \sin\chi \cos\chi
.
\end{eqnarray}
It is easy to see that case $< \alpha > = 1$ corresponds to 
"stationary"\  
or, better to say, equilibrium value of the inclination angle, 
when $< \frac{d\chi }{dt } > = 0$
\cite{Goldreich1970,GurevichIstomin2007,IstominShabanova2007}.
In this case the average speed of angular velocity decrease
also does not depend on the inclination angle: 
\begin{equation}
< \frac{ d\Omega }{ dt } > = - \frac{\Omega}{\widetilde{\tau}} 
.
\end{equation}
The last equation may correspond to the results of 
\cite{Biryukov2006,Biryukov2007}
where it is shown that an average braking index 
of many pulsars
is close to $n \approx 5$.
It is worth to note that if 
the parameter $< \alpha >$ decreases
with increasing angle $\chi$, 
this equilibrium state is stable,
see \cite{Tsygan2009} for details.

\section{Nondipolar magnetic field}
It is widely accepted that besides large-scale dipolar magnetic field, 
small scale nondipolar magnetic field may exist near neutron star surface.
This field has a spatial scale about $\ell \sim (0.3-3) km$ 
and rapidly falls to zero 
when the distance from the neutron star surface increases
to become fully negligible at the light cylinder.
This allows to suppose that this component does not exert 
any direct influence on  magnetic dipole braking,
that appears to occur near the light cylinder.
It seems that the same reasons are applicable to 
the direct influence of
the small scale component on current losses,
see for example \cite{Tsygan2009}.
However, the existence of the small scale component 
may lead to drastic changes 
of the electric current that flows through the inner gaps of the pulsar
\cite{Shibata1991_DC_Circuit}.
Thus, the indirect influence of the nondipolar field 
on current losses and torque $\vec{K}_{cur}$ 
is supposed to play a decisive role in pulsar braking.

\begin{figure}
  \includegraphics[width=0.6\hsize]{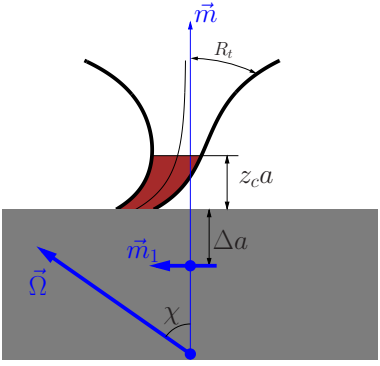}  
\caption{
  The position of small dipole $\vec{m}_{1}$ is shown
  in the case of $\gamma=0$ and $\phi = 0$.
  The lower gray square is neutron star
  and upper dark area is pulsar diode.
  The pulsar diode is situated on star surface $\eta_{0}=1$,
  $R_{t}$ is the radius of pulsar tube and
  $z_{c} a$ is the height of the diode.
  }
\label{pict_ns_nf}
\end{figure}

\begin{figure}
  \includegraphics[width=0.7\hsize]{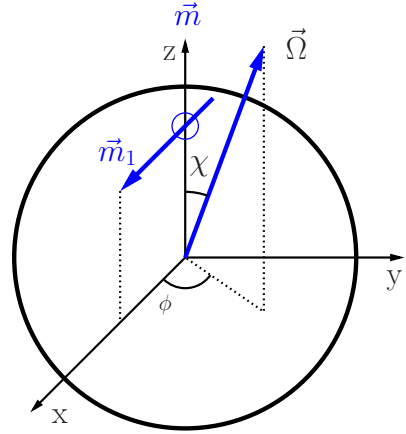}  
\caption{
  The orientation of 
  vectors $\vec{\Omega}$, $\vec{m}$ and $\vec{m}_{1}$
  in the case of $\gamma = 0$. 
  }
\label{pict_ns_star}
\end{figure}

The influence of small scale magnetic field upon electric currents 
was considered in many papers, 
c.f. \cite{Arons1979,
Gil2001,Gil2003_Drifting_and_inner,Gil2005_Interrelation,Gil2006_Formation}.
In order to estimate its influence on current losses 
we will use a simple model, proposed in \cite{PalshinPreprint},
see also \cite{Kantor2003,Tsygan2009}.
According to this model the inner gaps 
fully occupy the pulsar tubes cross sections 
and are situated close to the neutron star surface.
It is assumed that magnetic field strength on the star surface 
is not large enough 
to prevent free emission of electrons,
so when pulsar diode is placed onto star surface it will be operate
in electron charge limited steady flow regime.
In the neighbourhood of the inner gap small scale magnetic field 
will be modeled by the field of a small magnetic dipole $\vec{m}_{1}$,
which is embedded into neutron star crust 
under  magnetic (dipolar) pole at the depth $\Delta \cdot a$
(see fig. \ref{pict_ns_nf} and fig. \ref{pict_ns_star}).
We assume that $\Delta \approx 1 / 10$ and 
that $\vec{m}_{1}$ is perpendicular to the main dipole momentum $\vec{m}$.
So the vector $\vec{m}_{1}$ may be written in the form
\begin{equation}
\vec{m}_{1} =  2 m 
               \nu \Delta^{3} \left(
                                     \vec{e}_{x} \cos\gamma
                                    +\vec{e}_{y} \sin\gamma
                             \right)
,
\end{equation}
where $\nu = B_{1} / B_{0}$ is ratio of 
small scale magnetic field strength 
$B_{1} = m_{1} / (\Delta \cdot a)^{3} $
to the strength $B_{0}=2 m / a^{3}$ 
of the large scale dipolar magnetic field 
at the magnetic (dipolar) pole.
We also assume that the changes of values $\nu$, $\Delta$ and $\gamma$,
as well as, the value of magnetic momentum $m$,
are negligible, 
at least, over a period $T_{p}$ of neutron star precession.
A similar model has been proposed in \cite{Mitra2001},
where a small dipole $\vec{m}_{1}$ 
is placed not strictly under the magnetic pole 
and may have arbitrary direction.
The case of pure axisymmetrical small scale magnetic field 
has been investigated 
in \cite{Mitra2001,Mitra2002,Asseo2002,Tsygan2000}.
It is worth to note that the field of dipole $\vec{m}_{1}$ 
is able to describe small scale magnetic field 
only in vicinity of the pulsar diode.
We does not any intention to suppose 
that this dipole is able to describe small scale magnetic component 
over whole neutron star surface.
We rather suppose that the small scale field is close to 
a sum of $10^{2}-10^{3}$ such small dipoles 
and that we use only one of them, which is the closest to pulsar diode.

In the case of a thin pulsar tube $R_{t} \ll \Delta \cdot a$, 
where $R_{t}$ is its radius,
and at small altitudes 
$\eta = r / a \ll \kappa \sqrt{ c / \Omega a } \sim 10^{2}$
the Goldreich-Julian density (inside the pulsar tube) may be written as
\begin{equation}
\rho_{GJ} = - \frac{ \Omega B }{ 2\pi c } f(\eta)
.
\end{equation}
At $\nu \la 1 / 3 $ the function $f(\eta)$ may be approximated as 
\cite{PalshinPreprint}
\begin{eqnarray}
f(\eta) =   \frac{1}{ \sqrt{ 1 +\lambda^{2} } }
            \left(
                     \left( 1 - \frac{\kappa}{\eta^{3}} \right)
                     \cos\chi
            \right. 
\nonumber
\\
            \left.
                   + \lambda
                     \left( 1 +\frac{1}{2} \frac{\kappa}{\eta^{3}} \right)
                     \sin\chi\cos(\phi -\gamma)
            \right)
\label{feta_def}
,
\end{eqnarray}
where the coefficient $\kappa \approx 0.15$ describes 
general relativistic frame dragging \cite{Tsygan1992},
$B$ -- magnetic field strength,
$
\lambda = \nu\left( \Delta \eta \right)^{3}
             /
             \left(
                    \eta -1+\Delta
             \right)^{3} 
$.
In the case of $k=0$ the function $f(\eta)$ is equal to
the cosine of the angle between 
the direction of magnetic field $\vec{B}$ 
and angular velocity $\vec{\Omega}$.

In the case of a thin long inner gap,
when the radius $R_{t}$ of pulsar tube is small compared with the gap height
$z_{c} a$, see fig. \ref{pict_ns_nf},
the density $j$ of electric current flowing through the inner gap 
may be written as \cite{Tsygan2009}
\begin{equation}
j \approx j_{GJ}^{0} f(\eta_{0})
,
\end{equation}
where $\eta_{0}$ is the altitude where the inner gap begins
and $j_{GJ}^{0} = \frac{\Omega B}{ 2 \pi }$.
Following \cite{Tsygan2009} we assume that 
the electric potential $\Phi$ monotonically increases 
inside the pulsar diode 
and 
the diode is placed as close to the surface as possible.
Hence using expression (\ref{feta_def}) 
one may write at $\cos(\phi -\gamma )<0$
\begin{equation} 
j(\nu,\gamma) \approx 
j_{GJ}^{0} \cdot max\left( 0, f(1) \right)
\label{j_neg_cosgamma}  
\end{equation}
and in the case of $\cos(\phi -\gamma )>0$ 
\begin{equation}
j(\nu ,\gamma) \approx 
j_{GJ}^{0} \cdot min\left( f(\eta), 1 \leq \eta \leq +\infty \right)
\label{j_pos_cosgamma}
, 
\end{equation}
see \cite{Tsygan2009} for details.

Suppose that the small scale magnetic field near the northern inner gap 
may be described by a small dipole with parameters 
$\nu = \nu_{N}$ and $\gamma = \gamma_{N}$,
and the small scale magnetic field in a neighbourhood of southern inner gap
(and, of course, inside it) may be described 
by a small dipole with $\nu = \nu_{S}$ and $\gamma = \gamma_{S}$.
Then, the parameter $\alpha$ may be written as
\begin{equation}
\alpha = \frac{3}{4\cos\chi}
         \left(
                   \frac{ j( \nu_{N}, \gamma_{N}) }{ j_{GJ}^{0} }
                   \left( \frac{ S_{N}(\eta) }{ S_{0}(\eta) } \right)
                 + \frac{ j( \nu_{S}, \gamma_{S}) }{ j_{GJ}^{0} }
                   \left( \frac{ S_{S}(\eta) }{ S_{0}(\eta) } \right) 
         \right)
\label{alpha_two_pole_nondip} 
\end{equation}
where $S_{N}$ and $S_{S}$ is cross section areas of northern 
and southern pulsar tube.

\begin{figure}
   \includegraphics[width=0.90\hsize]{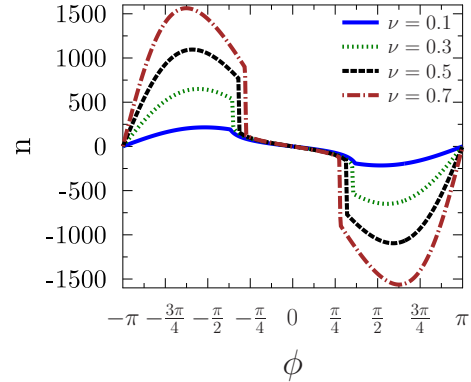}~
\caption{
   The dependence of braking index $n$ on precession phase $\phi$.
   The changing of phase $\phi$ from $-\pi$ to $\pi$ 
   corresponds to one precession period 
   and takes $T_{p}$ seconds.
   It is assumed that
   $S_{N}=S_{S}=S_{0}$,  
   $\nu_{N}=\nu_{S}=\nu$,
   $\gamma_{N}=\gamma_{S}=0$,
   $\chi = 30^{\circ}$,
   $P=1 {\mathrm s}$
   and the neutron star is spherical. 
  }
\label{pict_n_spher_pc_chi30_d0d00}
\end{figure}

\begin{figure}
   \includegraphics[width=0.90\hsize]{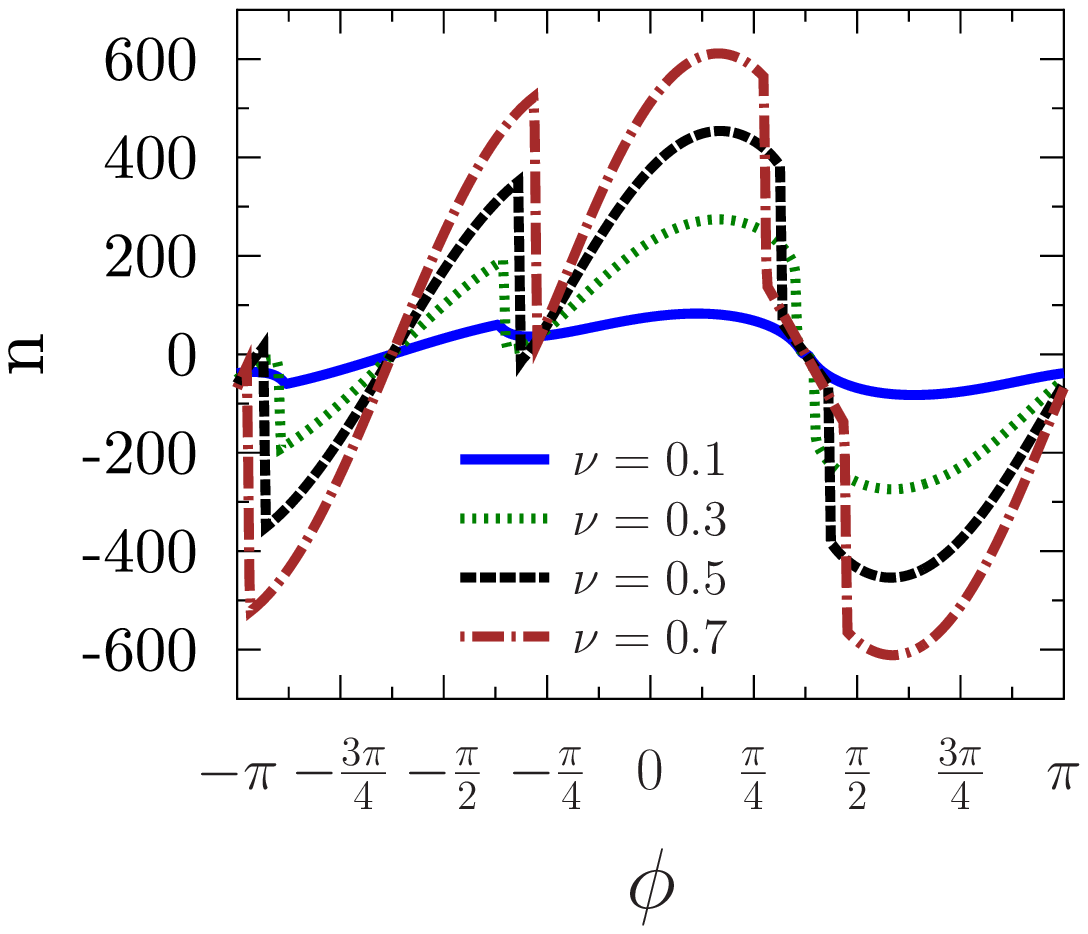}~ 
\caption{
   The same as fig. \ref{pict_n_spher_pc_chi30_d0d00},
   but $\gamma_{N}=0$ and $\gamma_{S}=\frac{3\pi}{4}$.
  }
\label{pict_n_spher_pc_chi30_d0d75}
\end{figure}

Let us firstly consider the most simple case when 
$\nu_{N}=\nu_{S}=\nu$ and $S_{N}=S_{S}=S_{0}$.
The values of braking index $n$ 
for angle $\chi = 30^{\circ}$ 
are shown on fig.
\ref{pict_n_spher_pc_chi30_d0d00}
and \ref{pict_n_spher_pc_chi30_d0d75}.
They have been calculated by the substitution of 
expressions (\ref{feta_def}) and (\ref{alpha_two_pole_nondip}) 
into equation (\ref{n_R_spher}).
It is easy to see that the values $\nu \approx 0.1$ 
mostly correspond to $n \sim 10-10^{2}$ and
for $\nu \sim 1$ 
the braking index may be as large as $\sim 10^{3}$.

\begin{figure}
   \includegraphics[width=0.90\hsize]{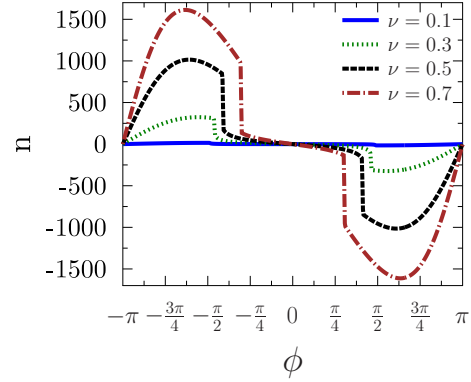}~
\caption{
   The dependence of braking index $n$ on precession phase $\phi$.
   The changing of phase $\phi$ from $-\pi$ to $\pi$ 
   corresponds to one precession period 
   and takes $T_{p}$ seconds.
   It is assumed that
   $S_{N}=S_{S}=S_{0}$,  
   $\nu_{N}=\nu_{S}=\nu$,
   $\gamma_{N}=\gamma_{S}=0$,
   $\chi = \chi_{eq}$,
   $P=1 {\mathrm s}$
   and the neutron star is spherical.
   Equilibrium angles $\chi_{eq}$ are equal to
   $\chi_{eq}\approx 81^{\circ}$ at $\nu=0.1$,
   $\chi_{eq}\approx 65^{\circ}$ at $\nu=0.3$,
   $\chi_{eq}\approx 49^{\circ}$ at $\nu=0.5$ and
   $\chi_{eq}\approx 33^{\circ}$ at $\nu=0.7$.
  }
\label{pict_n_spher_pc_equil_chi_d0d00}
\end{figure}

\begin{figure}
   \includegraphics[width=0.90\hsize]{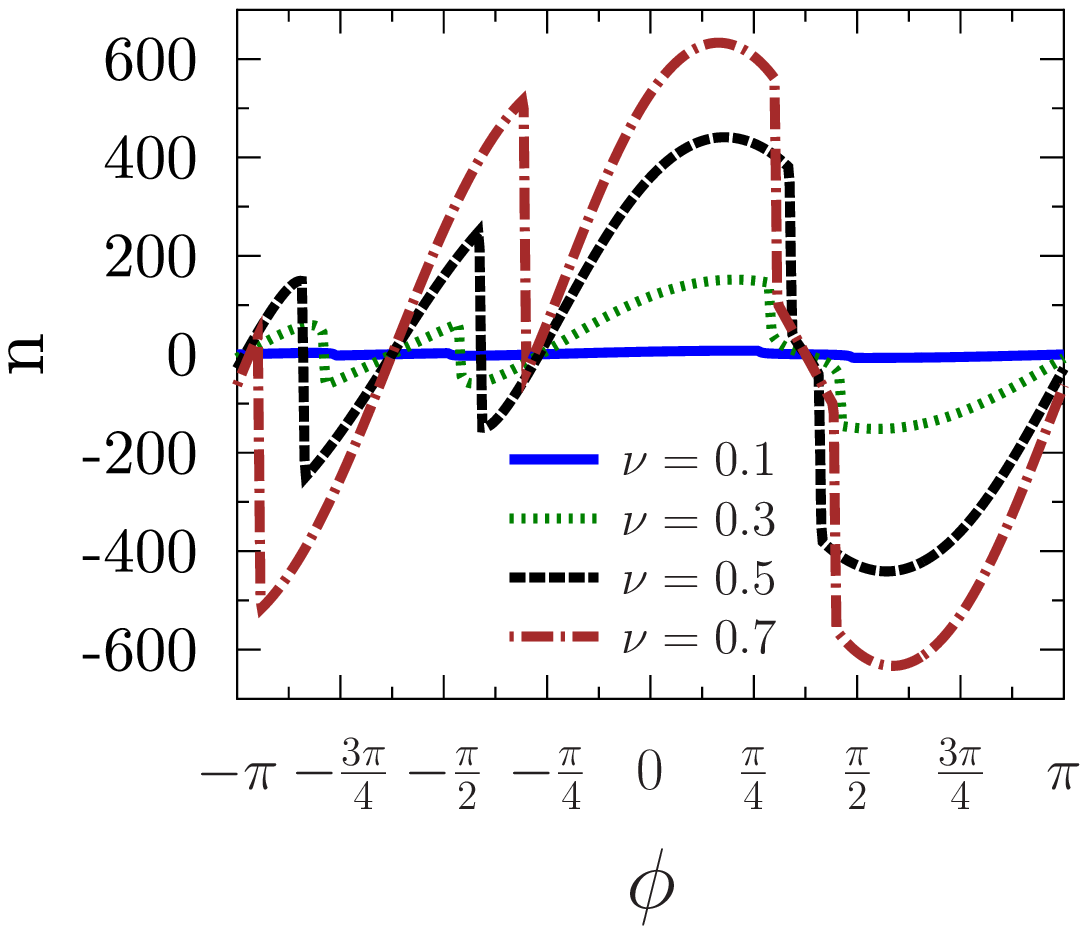}~
\caption{
   The same as fig. \ref{pict_n_spher_pc_equil_chi_d0d00},
   but $\gamma_{N}=0$ and $\gamma_{S}=\frac{3\pi}{4}$.
   Values of equilibrium angles $\chi_{eq}$ is the same.
  }
\label{pict_n_spher_pc_equil_chi_d0d75}
\end{figure}

On fig.
\ref{pict_n_spher_pc_equil_chi_d0d00}
and \ref{pict_n_spher_pc_equil_chi_d0d75}
it is shown the braking indices when
angle $\chi$ is not arbitrary 
but correspond to equilibrium value $\chi_{eq}$,
which is defined as 
$< \alpha >(\chi = \chi_{eq} ) = 1$.
It is easy to see that
in this case at $\nu=0.1$ the braking index
$n \sim \pm (10-20)$ and 
at $\nu = 0.7$ the braking index may achieve 
values as large as $n \sim 1.5 \cdot 10^{3}$.

In the case of weak small scale magnetic field 
$\nu \ll 1$ and $\nu \tan \chi < 1$ 
the electric current through the inner gap may be approximated as
\begin{equation}
j \approx j_{GJ}^{(0)} \cos\chi
                       \left( 1 - \kappa 
                                - \nu \tan \chi \cos(\phi - \gamma)
                                      \Theta( \cos(\phi - \gamma) )
                       \right)
\label{j_simple}
,
\end{equation}
where $\Theta(x)$ is the Heaviside function, 
$\Theta(x)=1$ at $x \geq 0$ and $\Theta(x)=0$ at $x<0$.
Consequently, parameter $\alpha$ is equal to 
\begin{equation}
\alpha = \frac{3}{4\cos\chi} 
                      \frac{1}{j_{GJ}^{0}}
                      \left(    j(\nu_{N},\gamma_{N} ) 
                              + j(\nu_{S},\gamma_{S} )
                      \right) 
\label{alpha_simple}
\end{equation}
The value of parameter $\alpha$ averaged over precession period $T_{p}$ 
may be written as
\begin{eqnarray}
< \alpha > & = & 
             \frac{3}{2} \left(
                            1 - \kappa
                           -\frac{\nu_{N}+\nu_{S}}{2\pi} \tan\chi
                         \right)
\label{av_alpha_simple}
.
\end{eqnarray}
Thus, the equilibrium angle $\chi$ 
is equal to
\begin{equation}
\chi_{eq} = {\mathrm arctan }
                  \left( 
                         \frac{2\pi}{\nu_{N}+\nu_{S}} 
                         (\frac{1}{3}-\kappa) 
                  \right)
\label{chi_equil_simple}
.
\end{equation}
At this angle the expression (\ref{alpha_simple}) may be rewritten as
\begin{eqnarray}
\alpha & = &
         \frac{3}{2} \left(
                             1 - \kappa
                     \right. 
                               -  
                             \frac{\pi}{3} ( 1 - 3\kappa )
                     \times
\nonumber
\\
\, & \times & 
                             \left(
                                    \frac{ \nu_{N} }{ \nu_{N} + \nu_{S} }
                                    \cos(\phi - \gamma_{N})
                                    \Theta( \cos(\phi - \gamma_{N}) )
                             \right. 
\nonumber
\\
\, & \, & +
                     \left.  
                             \left.
                                    \frac{ \nu_{S} }{ \nu_{N} + \nu_{S} }
                                    \cos(\phi - \gamma_{S})
                                    \Theta( \cos(\phi - \gamma_{S}) ) 
                             \right)
                     \right)
\label{alpha_equil_simple}
\end{eqnarray}
and in the case of $\nu_{N}=\nu_{S}$ and $\gamma_{N}=\gamma_{S}=0$ 
parameter $\alpha$ may be estimated as
\begin{equation}
\alpha \approx \frac{3}{2} \left(
                                 1 - \frac{1}{2} 
                                     \cos\phi \cdot \Theta(\cos\phi)
                           \right)
.
\end{equation}
This expression shows that at the equilibrium value of angle $\chi$ 
parameter $\alpha$ is changing 
from $\approx \frac{3}{4}$ to $ \approx\frac{3}{2}$
or, in other words, is changing in two times over the precession period.
Consequently, the braking index $n$ may be a crudely estimated as
$n \sim 2 \pi \widetilde{\tau} / T_{p}  \sim 10^{3} $.

Now let us discuss a more sophisticated case 
when the pulsar tube cross section $S_{t}$ depends on angle $\chi$.
We will use the dependence that was derived in \cite{Biggs1990}:
\begin{equation}
   S_{t}(\eta) \approx S_{0}(\eta) g^{2}(\chi) 
\end{equation} 
where 
\begin{equation}
g(\chi) =        \left(
                              \frac{ (1-\frac{\mu}{3})^{3} }{ 1 + \mu }
                 \right)^{1/4}
\end{equation}
and
$
\mu = 1 + \frac{1}{2}\cos^{2}\chi 
        - |\cos\chi | \sqrt{ 2 + \frac{1}{4} \cos^{2}{\chi} }
$.
When $\chi$ increases from $0$ to $\frac{\pi}{2}$ 
the value $\mu$ changes from $0$ to $1 / 3$.
At $\chi = 0^{\circ}$ the coefficient $g(\chi)$ is equal to $1$ 
and at $\chi = 90^{\circ}$ it is equal to
$
g\left( \frac{\pi}{2} \right) 
        =  \left(   \frac{4}{27}   \right)^{\frac{1}{4}}
                    \approx
                    0.620
$,
so the area of pulsar tube cross section decreases 
with increasing angle $\chi$.
Some other dependencies of pulsar tube areas on the angle $\chi$ 
is calculated, for example, in \cite{Rudak2004,Muslimov2009,Beskin_book}.
Such a dependence may be either decreasing or increasing, 
for example, in \cite{Beskin_book} it is shown 
that the area of a pulsar tube increases with $\chi$ 
from $1.59 S_{0}$ at $\chi=0^{\circ}$ 
upto $1.96 S_{0}$ at $\chi=90^{\circ}$.

\begin{figure}
   \includegraphics[width=0.90\hsize]{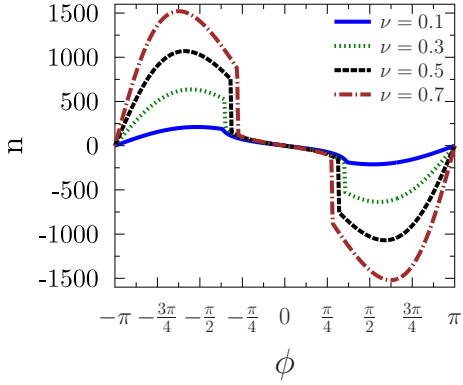} 
\caption{
   The same as fig. \ref{pict_n_spher_pc_chi30_d0d00},
   but $S_{N}=S_{S}=S_{0} g^{2}(\chi)$.
  }
\label{pict_n_spher_pc_biggs_chi30_d0d00}
\end{figure}

\begin{figure}
   \includegraphics[width=0.90\hsize]{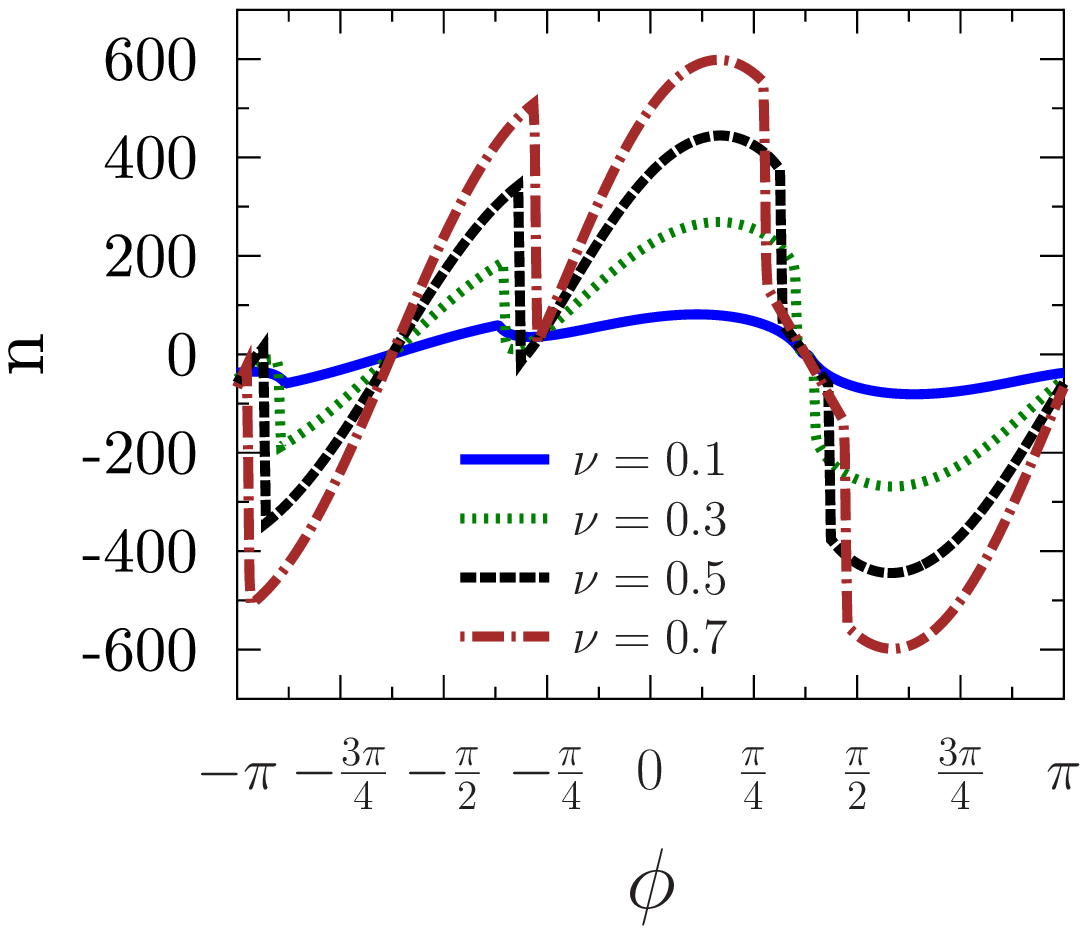} 
\caption{
   The same as fig. \ref{pict_n_spher_pc_biggs_chi30_d0d00},
   but $\gamma_{N}=0$ and $\gamma_{S}=\frac{3\pi}{4}$.
  }
\label{pict_n_spher_pc_biggs_chi30_d0d75}
\end{figure}

\begin{figure}
   \includegraphics[width=0.90\hsize]{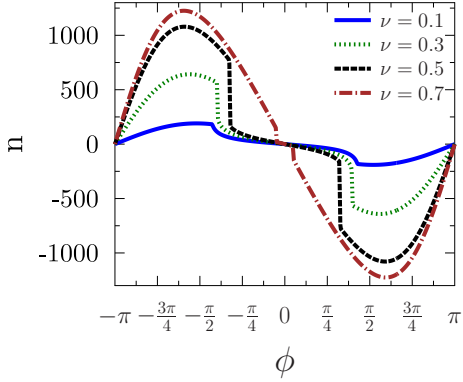} 
\caption{
   The same as fig. \ref{pict_n_spher_pc_equil_chi_d0d00},
   but $S_{N}=S_{S}=S_{0} g^{2}(\chi)$.
   Equilibrium angles $\chi_{eq}$ are equal to
   $\chi_{eq}\approx 46^{\circ}$ at $\nu=0.1$,
   $\chi_{eq}\approx 39^{\circ}$ at $\nu=0.3$,
   $\chi_{eq}\approx 30^{\circ}$ at $\nu=0.5$ and
   $\chi_{eq}\approx 21^{\circ}$ at $\nu=0.7$.
  }
\label{pict_n_spher_pc_biggs_equil_chi_d0d00}
\end{figure}

\begin{figure}
   \includegraphics[width=0.90\hsize]{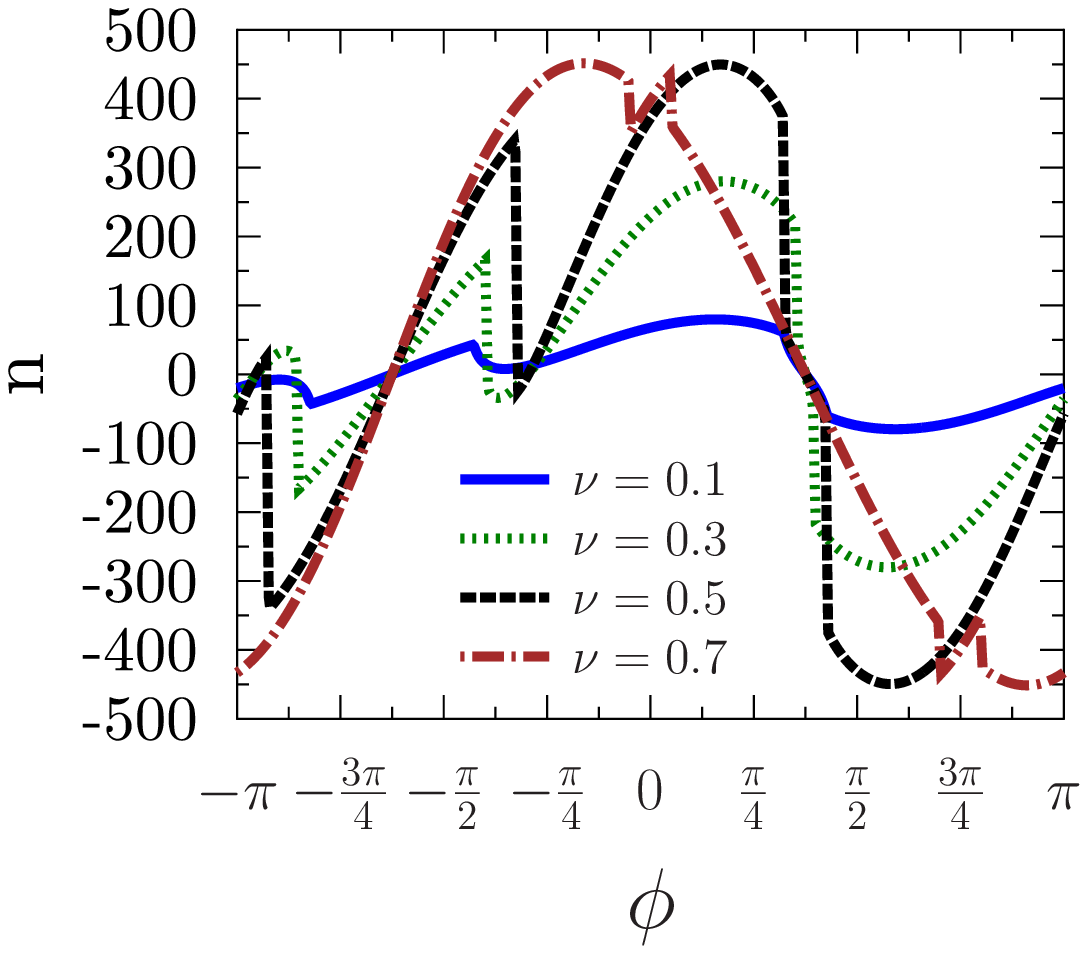}  
\caption{
   The same as fig. \ref{pict_n_spher_pc_biggs_equil_chi_d0d00},
   but $\gamma_{N}=0$ and $\gamma_{S}=\frac{3\pi}{4}$.
   Values of equilibrium angles $\chi_{eq}$ is the same.
  }
\label{pict_n_spher_pc_biggs_equil_chi_d0d75}
\end{figure}

In the case of $\nu_{N}=\nu_{S}=\nu$ and
$S_{N}=S_{S}=S_{0} g^{2}(\chi)$ 
the resulting braking indices $n$ 
for angle $\chi = 30^{\circ}$ 
are shown on fig.
\ref{pict_n_spher_pc_biggs_chi30_d0d00}
and \ref{pict_n_spher_pc_biggs_chi30_d0d75}.
Braking indices at equilibrium values of the angle $\chi$
are shown on fig.
\ref{pict_n_spher_pc_biggs_equil_chi_d0d00} 
and \ref{pict_n_spher_pc_biggs_equil_chi_d0d75}.
%

\section{An axisymmetric case}
In this section we will suppose that a neutron star is 
an absolutely rigid axisymmetric body 
and that its symmetry axis coincides with 
dipolar magnetic momentum $\vec{m}$.
The rotation momentum of the star may be written as
\begin{equation}
\vec{M}  = \sum_{\beta =x}^{z} I_{\beta} \cdot
                              \vec{e}_{\beta} \cdot
                              \left( \vec{e}_{\beta} \cdot \vec{\Omega} \right)
         =   I_{\perp} \vec{\Omega} 
           - \Delta I  \vec{e}_{m} \left( \vec{e}_{m} \cdot \vec{\Omega} \right) 
,
\end{equation}
where $I_{x}=I_{y}=I_{\perp}$ and $I_{z}=I_{||} = I_{\perp} -\Delta I$ 
are momenta of inertia of the star.
Hence, the equation (\ref{eqn_M_balance}) of momentum loss takes the form:
\begin{equation}
I_{\perp} \frac{ d\vec{\Omega} }{ dt } = \vec{K}_{eff} 
    = \vec{K} + \Delta I (\vec{\Omega} \cdot \vec{e}_{m} )
                         [ \vec{\Omega} \times \vec{e}_{m} ]
\label{Eqn_vec_domega_dI} 
,
\end{equation}
where $\vec{K}_{eff}$ is an effective rotation torque 
acting on the star
\begin{eqnarray}
\vec{K}_{eff} = K_{0} \left( \vec{e}_{m} 
                             \left(  1 - \alpha( \Omega,\chi , \phi )  \right)
                             \cos\chi
                           - \vec{e}_{\Omega}
                      \right.
\nonumber
\\
                      \left.
                           + R_{eff} 
                             [ \vec{e}_{\Omega} \times \vec{e}_{m} ]
                      \right)
\label{Keff_def} 
.
\end{eqnarray}
Here we neglect the small term 
$\Delta I \vec{e}_{m} \left( \vec{e}_{m} \cdot \dot{\vec{\Omega}} \right)
= ( \Delta I / I_{z} ) \vec{e}_{m} 
                       \left( \vec{e}_{m} \cdot \vec{K} \right)
$ 
and introduce the coefficient
\begin{eqnarray}
R_{eff} & = & 
           ( 
            \frac{ \Delta I \Omega^{2} }{ K_{0} } 
          + \frac{3 \xi}{2} \left( \frac{ c }{ \Omega a } \right)
           ) \cos\chi
\nonumber
\\
\,      & = &  
            \frac{3}{2} \left( \frac{c}{\Omega a} \right)
            \left( \xi  +  \frac{ \Delta I c^{2} a }{ m^{2} }   \right)
            \cos\chi 
\nonumber
\\
\, & \approx &
     3 \cdot 10^{4} \left( \frac{P}{ 1{\mathrm s} } \right)
                    \left( 
                           \frac{\xi}{4}  + 
                           \frac{\Delta I_{33}}{B_{12}^{2}}  
                    \right) 
                    \cos\chi
\label{Reff_def}
,  
\end{eqnarray}
where $\Delta I_{33} = \Delta I / 10^{33} {\mathrm g \, cm^{2}} $.

%
It is easy to see that the only difference between equations
(\ref{Eqn_vec_domega_spher}) and (\ref{Eqn_vec_domega_dI}) 
is the replacing of momentum of inertia $I$ by its component $I_{\perp}$
and the replacing of the coefficient $R_{dip}$ 
by the coefficient $R_{eff}$.
Consequently, upto these two exchanges 
all the formulas of previous section remain applicable 
to the axisymmetrical neutron star.
Particularly, the braking index $n$ may be estimated as
\begin{equation}
n \approx 
           R_{eff} \cdot
                   \frac{\partial \alpha }{ \partial \phi }
                   \cdot
           \frac{   
                  \cos^{2}\chi 
                }
                {
                   \left( \sin^{2}\chi  + \alpha \cos^{2}\chi \right)^{2}
                } 
  \sim R_{eff} \cdot \frac{\partial \alpha }{ \partial \phi }
\label{n_A_dI}  
.
\end{equation}
Again, supposing that $\alpha \sim 1$, $\cos\chi \sim \sin\chi \sim 1$,
one can crudely estimate the braking index as
\begin{eqnarray}
n & \sim & R_{eff} \sin\phi 
\nonumber
\\
\, & \sim &
   3 \cdot 10^{4} \left( \frac{P}{ 1\mbox{s} } \right)
                    \left( 
                           \frac{\xi}{4}  + 
                           \left( \frac{ \Delta I }{ 10^{-12} I } \right)
                           \frac{ I_{45} }{ B_{12}^{2} }
                    \right)  
   \sin\phi
\nonumber
\\
&  \sim & 
        \frac{ \tau }{ 250 \mbox{year} }                       
                       \left( \frac{ \Delta I }{ 10^{-11} I } \right)
                       I_{45}
                       \left(  \frac{ 1s }{ P } \right)
                       \sin\phi 
\label{est_n_at_DI}
,
\end{eqnarray}
where $\tau = P / (2\dot{P})$ is characteristic pulsar age
and $I \approx I_{\perp}$ 
is moment of inertia of a undeformed (spherical) neutron star,
$I_{45} = I / 10^{45} {\mathrm g \, cm^{2}}$.

\begin{figure}
   \includegraphics[width=0.90\hsize]{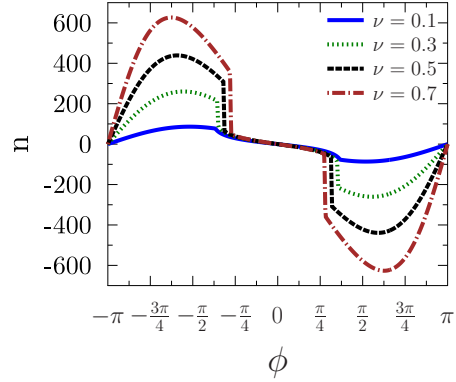}
\caption{
   The dependence of braking index $n$ on precession phase $\phi$.
   The changing of phase $\phi$ from $-\pi$ to $\pi$ 
   corresponds to one precession period 
   and takes $T_{p}$ seconds.
   It is assumed that
   $S_{N}=S_{S}=S_{0}$,  
   $\nu_{N}=\nu_{S}=\nu$,
   $\gamma_{N}=\gamma_{S}=0$,
   $\chi = 30^{\circ}$,
   $P=1 {\mathrm s}$,
   $B_{0} = 10^{12} G$
   and $\Delta I = - 10^{-13} I$.
  }
\label{pict_n_DIm1d0e32_pc_chi30_d0d00}
\end{figure}

\begin{figure}
   \includegraphics[width=0.90\hsize]{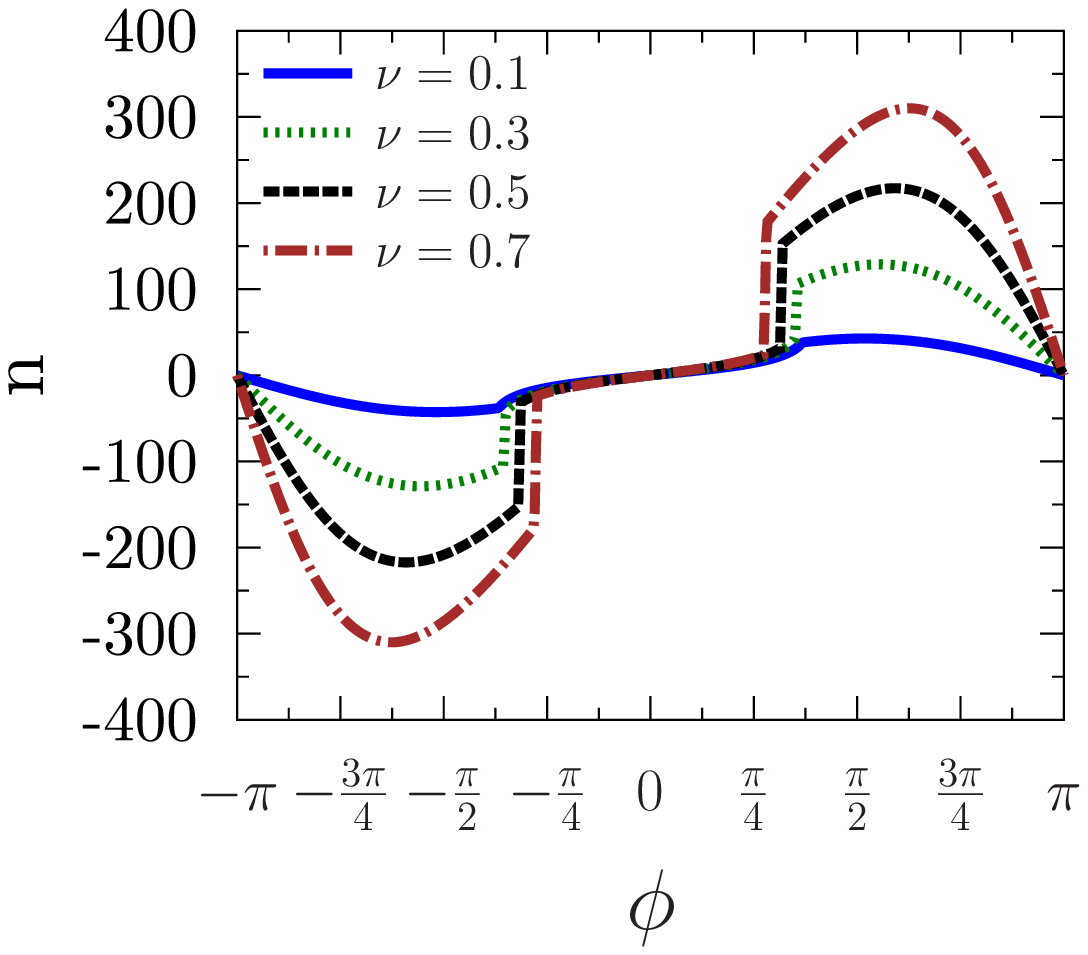}
\caption{
   The same as fig. \ref{pict_n_DIm1d0e32_pc_chi30_d0d00},
   but $\Delta I = - 2 \cdot 10^{-13} I$.
  }
\label{pict_n_DIm2d0e32_pc_chi30_d0d00}
\end{figure}

\begin{figure}
   \includegraphics[width=0.90\hsize]{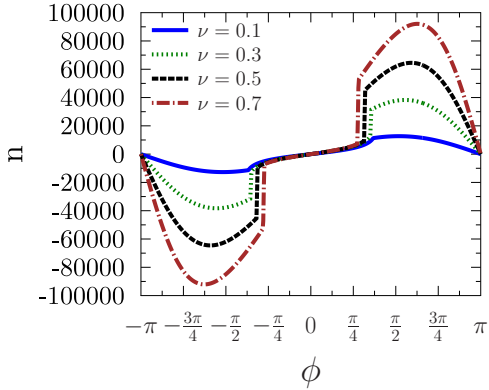}
\caption{
   The same as fig. \ref{pict_n_DIm1d0e32_pc_chi30_d0d00},
   but $\Delta I = - 1 \cdot 10^{-11} I$.
  }
\label{pict_n_DIm1d0e34_pc_chi30_d0d00}
\end{figure}

Some examples of dependence of braking index $n$ 
on the precession phase $\phi$ 
are shown on fig. 
\ref{pict_n_DIm1d0e32_pc_chi30_d0d00}-\ref{pict_n_DIm1d0e34_pc_chi30_d0d00}.
This dependence may be found by calculating the braking index 
$n_{spher}$ for a spherically symmetrical neutron star 
and then multiplying it by $R_{eff} / R_{dip}$:
\begin{equation}
n = \frac{ R_{eff} }{ R_{dip} } n_{spher}
.
\end{equation}

The axisymmetrical deformations of the neutron stars may be 
caused by internal magnetic field
that resides inside neutron star crust or inside its core
\cite{Goldreich1970}.
The deformation caused by internal magnetic field may be estimated as
\begin{equation}
   \frac{ \Delta I }{ I } \sim - \zeta \frac{ B_{in}^{2} a^{4} }{ G M^{2} }
,
\label{est_DI_at_Bin}
\end{equation} 
where $B_{in}$ is the strength of the internal magnetic field 
and $M$ is the mass of the star.
The coefficient $\zeta$ depends on magnetic field profile.
In the case of dipolar magnetic field 
$\zeta$  may be estimated as $\zeta = 25/8$ \cite{Ferraro1954},
the same value is used in \cite{Goldreich1970}.
For other configurations it may be, for example, 
as small as $\zeta = 1 / 18$
\cite{Glampedakis2008}.
If we assume that $B_{in} = B_{0} = 2 m / a^{3} $ then 
the coefficient $R_{eff}$ can be estimated as
\begin{eqnarray}
R_{eff} & \approx &
                 \frac{3}{2} \left( \frac{c}{\Omega a} \right)
                 \left(
                         \xi 
                       - 8 \zeta \frac{a}{r_{g}} \frac{ I }{ M a^{2} }
                 \right)
                 \cos\chi
\nonumber
\\
\, & \approx &
       3 \cdot 10^{4} \left( \frac{ 1s }{ P } \right)
         \left(
                 \xi - 12 \zeta \frac{a}{3r_{g}}  
                          I_{45}
         \right)
         \cos\chi
\label{est_Reff_at_Bin}
,
\end{eqnarray}
where $r_{g} = 2 G M / c^{2}$ is gravitational radius of the neutron star.
Firstly it is easy to see that 
in the case of $\zeta \sim 1 / 18$ 
the braking indices does change significantly.
Secondly if $\zeta \ga 1$ then 
the braking indices will be $\sim 20 \cdot \zeta$ times more
than in the spherically-symmetrical case 
and, particularly, 
at value $\zeta = 25 / 8$ used by \cite{Goldreich1970}
the braking index could become as large as $n \sim 10^{4}-10^{5}$.

In some cases the neutron star interiors may be superconductive.
If the neutron star matter is a superconductor of the second type, 
the internal magnetic field is able to form magnetic flux tubes.
In this case the coefficient $\zeta$ increase substantially
$\zeta \sim B_{fl} / B_{in} \sim 10^{3}$,
where $B_{fl}\sim 10^{15} {\mathrm G}$ is 
the strength of magnetic field inside magnetic flux tubes 
and $B_{in}$ is average strength of internal magnetic field
\cite{Wasserman2008}.
This leads to $B_{fl} / B_{in} \sim 10^{3}$ time larger 
deformation of the star 
and, consequently, to very short precession period like
\begin{equation}
  P_{p} \sim \frac{ 50 }{ \cos\chi } 
             \left( \frac{ P }{ 1s } \right)
             \mbox{ years }
.
\end{equation}
Hence, the coefficient $R_{eff}$ increases significantly
and leads to the increasing of braking index, 
that can reach $n \sim 10^{7}-10^{8}$.
As the absolute majority of normal isolated radiopulsars 
do not have such large braking indices 
it may be possible to conclude that 
there is no superconductivity of the second type in neutron star interiors.

\begin{figure}
  \includegraphics[width=0.6\hsize,angle=270]{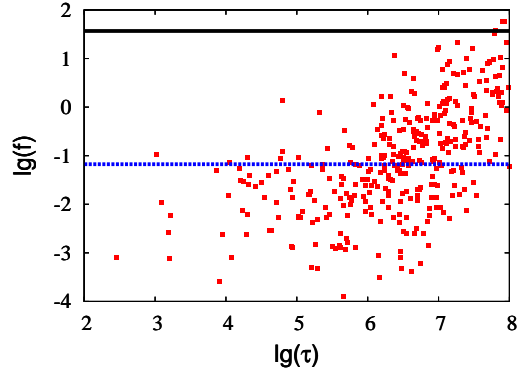}  
\caption{
  Dependence of value
  $f = \frac{2}{3} n \left( \frac{\Omega a}{ c } \right)$
  on the characteristic pulsar age 
  $\tau = \frac{ P }{ 2\dot P }$
  for various pulsars.
  The solid line corresponds to 
  value $f$ determined by (\ref{est_n_at_DI}) and (\ref{est_DI_at_Bin})
  $\zeta = 25 / 8$ 
  and the dashed line corresponds to $\zeta = 1 / 18$.
  Pulsar data is taken from \protect\cite{ATNF_catalog}.
  }
\label{pict_f_tau}
\end{figure}

The dependence of value
\begin{equation}
f = \frac{2}{3} n \left( \frac{\Omega a}{c} \right)
\end{equation} 
on characteristic pulsar age $\tau$ 
is shown on fig. \ref{pict_f_tau}.
The pulsar data is taken from \cite{ATNF_catalog}.
If the estimation (\ref{est_n_at_DI}) of braking index 
and the estimation (\ref{est_Reff_at_Bin}) of neutron star deformation
were valid 
then value $f$ would be depend only on parameter $\zeta$.
The solid line shows the estimation of value $f$ at $\zeta= 25 / 8$
and dashed line corresponds to the case of $\zeta= 1 / 18$.
Because of the estimation (\ref{est_n_at_DI}) yields only 
the upper limit of possible values of braking indices 
the observation data are in good agreement with 
the case of $\zeta = 25 / 8$.
Although the majority of pulsars does not contradict 
to the case of $\zeta= 1 / 18$ too.

\begin{figure}
  \includegraphics[width=0.6\hsize,angle=270]{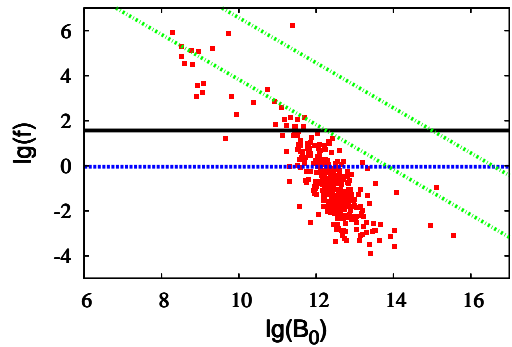}  
  \caption{
  Dependence of value
  $f = \frac{2}{3} n \left( \frac{\Omega a}{ c } \right)$
  on the dipolar magnetic field $B_{0}$ 
  for various pulsars.
  The solid line corresponds to 
  value $f$ determined by (\ref{est_n_at_DI}) and (\ref{est_DI_at_Bin})
  $\zeta = 25 / 8$ 
  and the dashed line corresponds to $\zeta = 1 / 18$. 
  The two dot-dashed lines correspond to the case of
  $\zeta = \frac{25}{8} ( B_{fl} / B_{0})$
  (upper line) and
  $\zeta = \frac{1}{18} ( B_{fl} / B_{0})$ 
  (lower line),
  $B_{fl} = 10^{15} {\mathrm G }$.
  Pulsar data is taken from \protect\cite{ATNF_catalog}.
  }
\label{pict_f_B}
\end{figure}

On fig. \ref{pict_f_B} it is shown the dependence of value $f$ 
on dipolar magnetic field $B_{0}$.
The pulsar data is taken from \cite{ATNF_catalog}.
Tho horizontal lines correspond to constant values of parameter $\zeta$.
Solid line corresponds to $\zeta = 25 / 8$ 
and dashed line corresponds to $\zeta = 1 / 18$.
The two dot-dashed lines correspond to the case of
$\zeta = (25 / 8) \cdot ( B_{fl} / B_{0})$
(upper line) and
$\zeta = ( 1 / 18) \cdot ( B_{fl} / B_{0})$ 
(lower line),
$B_{fl}= 10^{15} {\mathrm G}$.
The last two lines correspond to the case of 
neutron star deformations caused by magnetic flux tubes and 
the average internal magnetic field $B_{in}$ is equal to $B_{0}$.

\begin{figure}
  \includegraphics[width=0.6\hsize,angle=270]{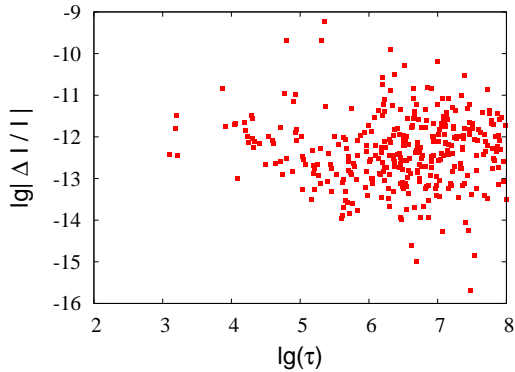}  
  \caption{
  The estimation of a lower limit of neutron star deformation $\Delta I / I$ 
  by braking indices for various pulsars.
  Value $\Delta I / I $ is estimated as
  $\frac{ \Delta I }{ I } \approx 
          \frac{2}{3} n \left( \frac{ \Omega a }{ c } \right)
                      \frac{ m^{2} }{ I a c^{2} }
  $.
  Pulsar data is taken from \protect\cite{ATNF_catalog}.
  }
\label{pict_DI_tau}
\end{figure}

The formula (\ref{est_n_at_DI}) allows to estimate the lower limit of
neutron star deformation $\Delta I / I$.
The corresponding values are shown on fig. \ref{pict_DI_tau}.
It shows that braking indices of majority of pulsars 
may be explained by the existence of deformation
$\Delta I / I \sim 10^{-13}-10^{-11} $
that agrees with estimation of $\Delta I$ received by \cite{Goldreich1970}.
%

\section{Conclusion}
In this paper we present a some explanation of large values of 
braking indices of pulsars.
The proposed model is based on four main assumptions:
\begin{enumerate}
\item
The neutron stars have a small scale magnetic field.
The strength of this field must be enough large 
to curve pulsar tube
but enough small to allow free emission of electrons from star surface.
\item
There are inner gaps in pulsar tubes
and 
a electric current flowing across inner gaps 
depends on angle between small scale magnetic field 
and angular velocity $\vec{\Omega}$ of the star.
\item
Braking torque depends on current that flows across inner gaps.
\item
The neutron stars precess with periods like 
$10^{3}-10^{4}$ years.
\end{enumerate}
In this paper we also neglect the contribution of outer gaps.
If a some part of electric current flows through outer gaps 
then this part is determined only by outer gap electrodynamics
and, consequently, does not depend from small scale magnetic field.
It decreases the variation of current losses over the precession period 
and, consequently, leads to the decreasing of braking index.
In the case of young pulsars like Crab and Vela,
which have small braking indices,
we suppose that current losses is almost fully determined
by currents flowing through outer gaps 
and the contribution of inner gaps current is negligible.
In this paper it is also assumed that
magnetic dipole braking exists and does not depends on
electric current flowing across inner gaps.
This assumption is widely used, 
c.f. \cite{Beskin2006_Be_born,Xu2007,GurevichIstomin2007,IstominShabanova2007},
but, as mentioned in \cite{Beskin_book},
it must be treated with caution.
It is shown that in the case of force free magnetosphere 
and absence of electric current flowing along pulsar tube 
the orthogonal pulsar $\chi = \pi / 2$ does not slow down at all
\cite{Beskin1983,Beskin1984,Shibata1999}.
It gives the reason to suppose that
magnetic dipole braking does not exist 
or, at least, must be depend on the electric current
\cite{Beskin_book}.
In such case presented model can provide 
the only qualitative explanation of 
the existence of large braking indices.
And the quantitative estimations, of course, will strongly depend on 
relation between magnetic dipole braking and current losses torques.

The presence of a long period precession is 
the weakest point of the model.
At present the precession is discovered only at a few isolated neutron stars.
And these pulsars have the precession periods like
$T_{p} \sim 1-10$ years \cite{Link2007}.
The presence of pined superfluid in neutron star crust 
substantially increases precession speed \cite{Shaham1977}. 
The precession periods $T_{p}$ larger than $(10^{2}-10^{4}) P$ 
may exist only when vortices of superfluid can not been pinned 
anywhere in star, 
are able to move freely 
and do not coexist with magnetic flux tubes
\cite{Link2006_Incompatibility,Link2007}.
%
The small number of isolated pulsars with observed precession 
force us to exclude the triaxial precession 
and to assume that neutron star is axisymmetrical 
and symmetry axis coincide with magnetic dipole moment $\vec{m}$.
In this case in frame reference related to "rigid stars" 
vector $\vec{m}$ rotate with constant angular velocity 
$\vec{\omega}$. 
And because of $\vec{\omega} \approx \vec{\Omega}$ 
its trajectory coincides with the case of unprecessing star, see Appendix A.
Also, in order to prevent the observation of such precession
it is necessary to assume that pulsar tube structure does not precess too.
It particularly means that pulsar tube crossection is circular 
or depends on only vectors $\vec{\Omega}$ and $\vec{m}$ 
and does not depend from small scale magnetic field.
Also, it means that distributions 
of energy of primary electrons 
and pair multiplicity 
over pulsar tube crossection
are close to axisymmetrical.

\section*{Acknowledgments}
This work was supported by 
the Russian Foundation for Basic Research
(project 10-02-00327)
and the Program of the State Support 
for Leading Scientific Schools of the Russian Federation
(grant NSh-3769.2010.2).
We sincerely thank A.M. Krassilchtchikov for 
help with the preparation of the paper,
A.I. Chugunov for help in the calculations 
and D.G. Yakovlev, A.I. Chugunov, M.E. Gusakov, 
Yu.A. Shibanov, D.A. Zyuzin, A.A. Danilenko,
E.M. Kantor, V.D. Palshin and V.A. Urpin
for help, comments, and usefull discussions.

\appendix

\section[]{The precession of axisymmetrical star} 
Consider the neutron star rotation in an extreme case when 
it is possible to neglect 
the first and second terms in expression (\ref{Keff_def}) 
which contain vectors $\vec{e}_{m}$ and $\vec{e}_{\Omega}$
compared with the last term which contains 
$[ \vec{e}_{\Omega} \times \vec{e}_{m} ]$.
Particularly, it means that any rotation energy losses are neglected
and it is assumed that the angular velocity $\Omega$ and 
inclination angle $\chi$ do not change with time.
Consequently, this approximation is valid only over time scales
that are small compared with the characteristic time $\widetilde{\tau}$ 
or pulsar age $\tau$,
although this time scales may be comparable 
with or larger than the precession period $T_{p}$.
In this case the equation (\ref{Eqn_vec_domega_dI}) may be written as
\begin{equation}
\frac{ d\vec{\Omega} }{ dt } =
      \frac{ 1 }{ \widetilde{\tau} }
      R_{eff}
      [ \vec{\Omega} \times \vec{e}_{m} ] 
.
\end{equation}
With this equation it easy to obtain  
$d\Omega / dt =0$ and $ d\chi / dt = 0$ 
and, consequently, 
neither the characteristic time $\widetilde{\tau}$ 
nor the coefficient $R_{eff}$ are changing.

Let us introduce the vector 
$\vec{e}_{\omega} = \cos\beta \vec{e}_{\Omega} - \sin\beta \vec{e}_{m} $
and calculate its time derivative
\begin{eqnarray}
\frac{ d \vec{e}_{\omega} }{ dt } =
  & - &
  ( \sin\beta \vec{e}_{\Omega} + \cos\beta \vec{e}_{m} ) 
    \frac{ d\beta }{ dt } 
\nonumber
\\
& + &
    (   \frac{ R_{eff} }{ \widetilde{\tau} } \cos\beta 
      - \Omega \sin\beta 
    )
    [ \vec{e}_{\Omega} \times \vec{e}_{m} ]
= 0
.
\end{eqnarray}  
Thus, in order to make vector $\vec{e}_{\omega}$ constant in time,
it is enough to choose the angle $\beta$ as:
\begin{equation}
\beta = - {\mathrm arctan }
                \left( 
                      \frac{ R_{eff} }{ \Omega \widetilde{\tau} } 
                \right)
.
\end{equation}
In this case the first of equations (\ref{dot_e_m}) may be rewritten as
\begin{equation}
\frac{ d\vec{e}_{m} }{ dt } = 
     \frac{ \Omega }{ \cos\beta } 
     [ \vec{e}_{\omega} \times \vec{e}_{m} ]
.
\end{equation} 
It means that $\vec{e}_{m}$ and, consequently, $\vec{m}$ 
just rotate with the constant angular velocity 
$\vec{\omega} = \vec{e}_{\omega}  \Omega / \cos\beta $ 
around the constant vector $\vec{\omega}$,
see for example \cite{Link2003}.
As 
$\vec{\Omega} = \vec{\omega} + \vec{e}_{m}\, \Omega\, \tan(\beta)  $, 
$\Omega$ would also rotate with the angular velocity $\vec{\omega}$
around the constant direction $\vec{e}_{\omega}$.
It is worth to note that the angle $\beta$ is very small:
\begin{eqnarray}
\beta & = & {\mathrm arctan}
                  \left(
                     \left(
                       \xi \frac{m^{2}}{I c^{2} a}
                      +\frac{\Delta I}{ I }
                     \right)
                     \cos\chi
                  \right)
\nonumber
\\
\, & \sim &
       10^{-12} \left(
                        \frac{\xi}{4}
                        \frac{ B_{12}^{2} }{ I_{45} }
                        \,
                      + \frac{\Delta I}{10^{-12} I}
                \right)
                \cos\chi
,
\end{eqnarray}
and, consequently, $\vec{\omega}$ is almost undistinguished 
from the star angular velocity $\vec{\Omega}$.
It seems that such a rotation substantially increases the difficulty of 
direct observations of pulsar precession.
\begin{figure}
   \includegraphics[width=0.6\hsize]{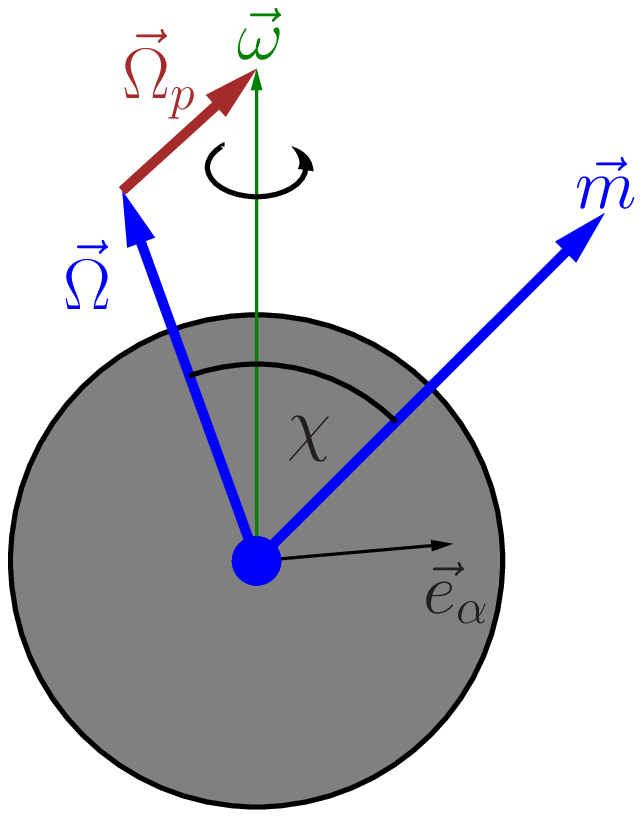}
\caption{
      The rotation of vector $\vec{e}_{\alpha}$ with angular 
      velocity $\vec{\Omega}_{p}$.
    }
\label{pict_Omegap}
\end{figure} 
It is easy to see, 
that in the reference frame $K_{\omega}$, 
rotating with the angular velocity $\vec{\omega}$, 
vectors $\vec{e}_{m}$ and $\vec{e}_{\Omega}$ are constant in time 
and the star rotates around vector $\vec{e}_{m}$ 
with the angular velocity $(-\vec{\Omega}_{p})$
\cite{Link2003}, 
see fig. \ref{pict_Omegap}, where
\begin{equation}
\vec{\Omega}_{p} = - \vec{e}_{m}\, \Omega\, \tan(\beta)
\end{equation}
Take, for example, any vector $\vec{e}_{\alpha}$ rotating 
together with the star as,
see fig. \ref{pict_Omegap}:
\begin{equation}
\frac{ d\vec{e}_{\alpha} }{ dt } =
    [ \vec{\Omega} \times \vec{e}_{\alpha} ]
\end{equation}
Hence in the reference frame $K_{\omega}$ it is possible to write
\begin{eqnarray}
    \frac{ d \vec{e}_{\alpha} }{ dt } & = &
             [ \vec{\Omega} \times \vec{e}_{\alpha} ]
            -[ \vec{\omega}     \times \vec{e}_{\alpha} ]     =
\nonumber
\\  
    \, & = &
           - [ \vec{\Omega}_{p} \times \vec{e}_{\alpha} ]
\nonumber
\end{eqnarray}  
Because of radio radiation may be generated at a some distance from 
the center of pulsar tube,
the precession may be manifested as subpulses drift 
or variation of pulse profile
\cite{Asseo2002,Gil2006_Mimicking}.
In order to exclude such manifestations it is necessary to assume 
that average distribution of radio sources in pulsar tube 
is axisymetrical.
Also, it needs to assume that pulsar tube crossection is circular 
or, at least, its profile depend on vectors $\vec{\Omega}$ and $\vec{m}$ 
and does not depend on small scale magnetic field.
The model of pulsar tube presented in \cite{Biggs1990}
satisfies this criteria.

\bsp

\label{lastpage}

\end{document}